# High-pressure self-flux growth and characterization of Li-deficient Li$_{0.95}$FeAs single crystals

N. D. Zhigadlo

*CrystMat Company, CH-8037 Zurich, Switzerland*

**Abstract**

Single crystals of LiFeAs were successfully grown by a self-flux method using the cubic anvil high-pressure and high-temperature technique. The reaction took place in a closed boron nitride crucible at a pressure of 1 GPa and a temperature of 1050 °C. The grown crystals, retrieved from solidified lump, exhibit plate-like shapes with sizes up to 0.7 × 0.7 × 0.15 mm$^3$. Single-crystal x-ray diffraction refinement confirmed the high structural perfection of the grown crystals [space group *P*4/*nmm*, No 129, $Z = 2$, $a = 3.77370(10)$ Å, $c = 6.3468(6)$ Å, and $V = 90.384(9)$ Å$^3$] with the presence of a small deficiency on the Li site. The refined chemical composition of the produced crystals is Li$_{0.95}$FeAs. Temperature-dependent magnetization measurements revealed bulk superconductivity with a superconducting transition $T_c = 16$ K. A comparative analysis of LiFeAs, Li$_{1-x}$FeAs, and Li$_{1-y}$Fe$_{1+y}$As systems reveal that superconductivity is less sensitive to the Li deficiency, although it may completely disappear in the presence of both Fe excess and Li deficiency.

*E-mail address:* nzhigadlo@gmail.com ; *URL*: http://crystmat.com

## 1. Introduction

The discovery of superconductivity in LaFeAs(O,F) with a superconducting transition temperature $T_c = 26$ K [1] has generated intense research activities on the iron-based superconductors. A large number of quaternary iron-oxypnictides *Ln*FeAsO (*Ln* = Ce, Pr, Nd and Sm), generally named 1111 systems, were found to be superconducting with $T_c$ up to 55 K [2]. Later on, also the ternary 122 system, *Ae*Fe$_2$As$_2$ (*Ae*: alkaline-earth), the 11 system, iron-chalcogenides Fe(Se,Te), as well as the ternary 111 system, *A*FeAs (*A* = Li, Na) were shown to belong to the class of iron-based superconductors [3-9].

Among the different *A*FeAs compounds, LiFeAs is one of the most intriguing with a superconducting critical temperature $T_c$ close to 18 K at ambient conditions [9]. Similar to the situation with MgB$_2$, this compound has been known since 1968 [10], but curiously no one tried to measure its superconducting properties. In that case, the era of Fe-based superconductors would have begun much earlier. It is interesting to note that LiFeAs behaves differently from many other iron-pnictide superconductors. First, the edge-sharing FeAs$_4$ tetrahedra are strongly compressed in the basal plane, since they need to accommodate the small Li$^+$ ion between the FeAs layers. Second, the material LiFeAs exhibits superconductivity already in its stoichiometric form, whereas other iron arsenide materials require doping to achieve a formal oxidation state of $Fe^{3+}$ or the application of hydrostatic- or chemical pressure to become superconductors. Third, unlike NaFeAs and other 122- and 1111 type iron arsenides, stoichiometric LiFeAs does not exhibit static magnetism.

LiFeAs crystallizes in the tetragonal $Cu_2Sb$/PbClF-type structure (*P4/nmm*) [9] and has a single FeAs layer sandwiched between two Li layers (Fig. 1). This implies that there is a unique opportunity to obtain a homogeneous Li terminating surface upon cleaving. This opportunity, based on the structural character, is very important for many spectroscopic techniques, including angle-resolved photoemission, scanning tunneling microscopy, and optical spectroscopy, since all of them are highly sensitive to the surface state [11,12]. The possibility of realizing a homogeneous surface and the structural simplicity make LiFeAs a good candidate to investigate the intrinsic properties of the Fe-based superconductors. Clearly, this requires the growth of clean and large single crystals of LiFeAs.

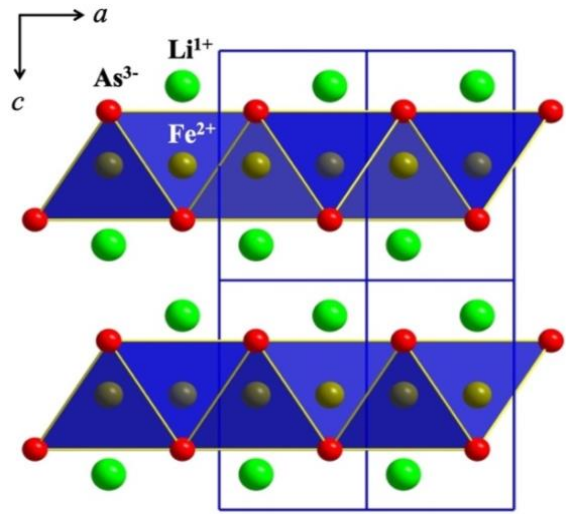

**Fig. 1.** Schematic representation of the LiFeAs lattice projected on the *ac* plane. Blue polyhedra indicate the $FeAs_4$ tetrahedra, separated by Li ions and stacked alternately along the *c* axis. The green circles are for Li atoms, the gray circles are for Fe, and the red circles are for As. Solid lines indicate the tetragonal unit cells with space group *P4/nmm*, No. 129 and lattice parameters $a = b = 3.7737$ Å, $c = 6.3468$ Å, $\alpha = \beta = \gamma = 90°$.

Several ways for producing polycrystalline and single-crystalline LiFeAs materials have been reported in the literature [5-9]. For example, a Sn-flux method, has been widely used to grow single crystals of various Fe-based superconductors [5,13-18]. Compared to other processes, the Sn-flux method has the advantage of the high solubility of most metallic elements, including Li, into the Sn flux, as well as the relatively wide growth-temperature window due to the low melting point of Sn, ~232 °C. However, it has the drawback of the possible inclusion of Sn into the crystal lattice, which may alter the intrinsic properties, as is the case of $BaFe_2As_2$ single crystals grown via the Sn-flux approach [14-18]. Tapp et al. [6] were the first who successfully grew LiFeAs single crystals by a solid-state reaction using stoichiometric amounts of Li, Fe, and As. Wang et al. [9] reported on the synthesis of LiFeAs under high pressure, which yielded nearly phase pure polycrystalline samples with maybe a slight off-stoichiometry. Two different methods of preparing polycrystalline LiFeAs were proposed by Pitcher et al. [7], starting from either Li and pre-reacted FeAs or from $Li_3As$, Fe, and FeAs. Tiny single crystals, a few tenths of a millimeter in size, were obtained by prolonged annealing at low temperature by Chu et al. [19]. Crystals with a size up to $6 \times 6 \times 3$ mm$^3$ can be grown in a sealed tungsten crucible using the Bridgman method [20]. However, their quality was questionable because a complete diamagnetic shielding could not be demonstrated and the resistivity curves were compatible with the presence of a foreign phase. A self-flux method has been shown to be effective in growing LiFeAs crystals using Li ingot and FeAs powder or the $Li_3As$, Fe, and As powders as a starting materials and a BN and Nb crucibles and quartz tubes. Because of the high melting temperature of the FeAs flux (~1030 °C) and the high volatility of Li ions, a metal crucible needs be welded under high Ar pressure [21,22]. Using the self-flux technique, Morozov et al. [23] obtained the largest crystals with lateral dimensions of $(12 \pm 6) \times (12 \pm 6) \times (0.3 \pm 0.05)$ mm$^3$ and terrace-like features. However, their overall perfection is unclear because most of their structural and physical measurements were conducted on much smaller sized crystals.

Overall, growing stoichiometric single crystals of LiFeAs has proven to be a fascinating challenge. In order to provide a suitable platform for the meaningful assessment of the structural, compositional, and physical properties of LiFeAs produced by diverse methods, as well as for a comparison with other pnictides families, high-quality single crystals with a well-defined stoichiometry are needed. The main objectives of the present study were to explore the application of a high-pressure, high-temperature approach for the growth of LiFeAs crystals without the addition of any external fluxes and to stabilize Li-deficient crystals. Compared to ambient conditions, high pressure can facilitate the creation of deficient structures by bringing atoms closer together and lowering vacancy formation energies. We have previously applied this approach to the growth of different types of $Ln$FeAsO ($Ln$ = Sm, Nd, Pr) superconductors [24-29]. These earlier successful experiments also suggest that this method might be used to grow single crystals of LiFeAs. By contrasting our findings with those already published for LiFeAs, $Li_{1-x}FeAs$, and $Li_{1-y}Fe_{1+y}As$ systems, we show that LiFeAs superconductivity is not very sensitive to Li deficiencies, yet it may entirely disappear in the presence of Fe excess. For this reason, compositional control is essential when comparing the characteristics of LiFeAs made by different methods.

## 2. Experimental details

In this article, we present the synthesis of LiFeAs using a semi-cylindrical multi-anvil module made by Rockland Research Corporation (USA). It works by squeezing six small inner tungsten carbide (WC) anvils with eight large outer steel dies, thus applying an amplified pressure on the innermost cubic cell. Figure 2a,b depicts the high-pressure sample-cell assembly utilized in the high-pressure growth as well as a basic overview of the cubic-anvil inner parts. A tubular graphite heater is installed in the bore of a pyrophyllite cube within the sample-cell assembly. Pyrophyllite tablets are loaded on top and bottom of the heater, together with a cylindrical boron nitride (BN) crucible in the middle. A current that passes through the steel components heats the graphite heater. The temperature of the sample is determined by the pre-calibrated relation between the applied electrical power and the measured temperature in the cell. A water-cooling mechanism prevents the WC anvils from overheating. More details regarding the high-pressure apparatus and the experimental setup can be found elsewhere [27, 29]. We have previously used the high-pressure method with great success to grow crystals of a variety of superconducting- [30-32] and magnetic materials [33-35], diamonds [36], cuprate oxides [37,38], pyrochlores [39,40], $Ln$Fe$Pn$O ($Ln$: lanthanide, $Pn$: pnictogen) oxypnictides [41,42], polymers [43], and a number of other compounds [44-47].

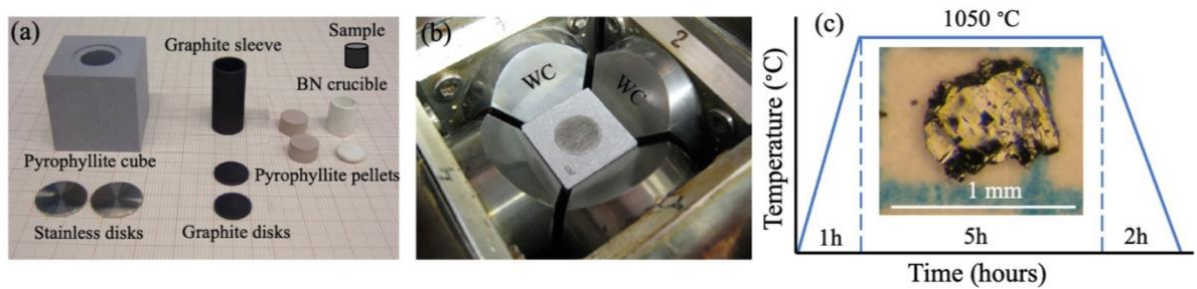

**Fig. 2.** Illustration of the sample cell assembly, high-pressure synthesis process, and example of obtained crystals. (a) The sample assembly for the cubic anvil high-pressure crystal growth: pylophyllite cube, stainless steel disks, graphite sleeve, graphite disks, pyrophyllite pellets, BN crucible and sample. (b) Top view on cubic experimental cell: pyrophyllite cube is placed in the core of cell between tungsten carbide (WC) anvils. (c) The furnace heat-treatment protocol used for the high-pressure self-flux growth of LiFeAs single crystals. The inset shows the optical microsope image of a LiFeAs single crystal mechanically extracted from solidified lump.

The LiFeAs crystals were grown by the high-pressure self-flux method. The starting materials, here, Li ribbons (purity 99.99%) and FeAs powder (purity 99.99%) were mixed according to the nominal formula LiFeAs. The FeAs precursor was synthesized in advance from high purity of Fe (99.99%) and As (99.99%) powders by sealing them into an evacuated tube and sintering it at 700 °C for several times until a single phase forms. Since some of the starting materials and the final product of LiFeAs are hygroscopic and react easily with oxygen, all the preparation and sample-handling procedures were carried out in a glove box containing extremely pure (99.9999%) argon gas with a combined $O_2$ and $H_2O$ content of less than 1 ppm.

The structural properties of LiFeAs single crystals were investigated at room temperature using a Bruker x-ray single-crystal diffractometer equipped with a charge-coupled device (CCD) detector and high intensity microfocus x-ray source (Mo K$_\alpha$ radiation, $\lambda = 0.71073$ Å). Data reduction and numerical absorption correction were performed by a direct method and refined on $F^2$, employing the SHELXL program [48]. The elemental analysis of the produced crystals was performed by means of Energy Dispersive X-ray Spectroscopy (EDX, Hitachi S-3000 N). A Quantum Design Magnetic Property

Measurement System (MPMS XL) with a Reciprocating Sample Option (RSO) was used to perform the temperature-dependent magnetic measurements. The crystalline samples were immobilized in gelatin capsules. Measurements were conducted in a DC field of 1 mT in the temperature range 5-25 K after cooling in zero applied field (zero-field-cooled: ZFC) and in the measuring field (field-cooled: FC).

### 3. Results and discussion

In a typical growth run, a pressure of 1 GPa was applied at room temperature. While keeping the pressure constant, the temperature was increased within 1 h to the maximum value of 1050 °C, kept constant for 5 h and then the cell was cooled to room temperature in 2 h (Fig. 2c). Throughout the synthesis the high pressure was kept constant and was released once the cell was cooled to room temperature. In addition to accelerating the reaction, the main benefits of applying high pressure and high temperature, are that they prevent Li from oxidizing or evaporating upon heating. The obtained crystals were rather fragile and prone to exfoliation. They cleave along the *ab* plane, in particular, between two layers of Li. The crystals are sensitive to air moisture, which complicates their handling and study. The insert in Fig. 2c shows an optical image of a typical crystal, demonstrating a size of about $0.7 \times 0.7 \times 0.15$ mm$^3$ and shiny and metallic-like surfaces. The layered crystal morphology mimics the layered type of the LiFeAs structure, consisting of FeAs and Li planes. A similar morphology was also observed in LiFeAs crystals grown by other methods [5,23]. Overall, it reflects the fact that the bonding between the adjacent FeAs layers, mediated via As-Li-As bonds, is weaker compared to the bonding within the FeAs layers. This causes a faster growth rate within the *ab* layers, while the growth along the *c* axis is slow, leading to the formation of thin and plate-like crystals. We note that almost all the FeAs-based crystals exhibit a similar appearance [18, 23-29], which is not surprising considering their crystallo-chemical closeness.

In our high-pressure growth experiment, the crystals grow mostly at the top- and bottom parts of the BN crucible, which are coolest areas in the high-pressure cell due to its directional temperature gradient. Further details on the temperature distribution in a high-pressure cell can be found in [29,30]. As can be seen from our heating protocol (Fig. 2c), we do not need to apply a specific cooling rate because the natural gradient is sufficient to initiate the growth. FeAs melts at ~1030 °C [49] and plays the role of flux, whereas the closed BN crucible at high pressure prevents Li from evaporating during the growth.

EDX analysis performed on several spots revealed that the Fe/As molar ratio is 1:1. Unfortunately, this method cannot provide an accurate estimate of the Li content in the crystal due to the unreliability of EDX in quantifying light elements. In fact, the stoichiometry of LiFeAs is an important issue, and often Li-deficient samples have been reported [50-53]. We note that, currently there is no convincing way to link these deficiencies with structural parameters. This is associated with the challenge of identifying the deficiency and, often, to the considerable compositional diversity. Li occupancies on Fe sites have been also reported [52,54]; all of this makes comparing samples made by various methods more difficult. Single-crystal data are therefore crucial.

The grown LiFeAs crystals crystallize in a tetragonal structure with space group *P4/nmm* (No. 129) and lattice parameters $a = 3.77370(10)$ Å, $c = 6.3468(6)$ Å and with the corresponding calculated cell volume of $V = 90.384(9)$ Å$^3$. The crystallographic data and the details of the structure refinement are summarized in Table 1. All crystallographic positions as well as the anisotropic displacement parameters are shown in Table 2. In the refinement we allowed the Li, Fe, and As site occupancies to vary. The structural analysis of our crystals indicates a 0.95 occupancy on the Li site, suggesting that the real composition is slightly different from the nominal and closer to Li$_{0.95}$FeAs.

**Table 1.**

Details of single-crystal X-ray diffraction data collection and crystal refinement results for Li$_{0.95}$FeAs.

| | |
|---|---|
| Identification code | Shelx |
| Empirical formula | Li$_{0.95}$FeAs |
| Temperature | 295(2) K |
| Wavelength | 0.71073 Å /MoK$\alpha$ |
| Crystal system | Tetragonal |
| Space group | *P4/nmm* |
| Z | 2 |
| Unit cell dimensions | $a = 3.77370(10)$ Å, $c = 6.3468(6)$ Å |
| Cell volume | 90.384(9) Å$^3$ |
| Density (calculated) | 5.047 g/cm$^3$ |
| Absorption correction type | analytical |
| Absorption coefficient | 25.917 mm$^{-1}$ |
| $F(000)$ | 124 |
| Crystal size | 0.132 × 0.72 × 0.47 mm$^3$ |
| Theta range for data collection | 6.29 to 45.64 deg. |
| Index ranges | $-6 \leq h \leq 5$, $-7 \leq k \leq 3$, $-12 \leq l \leq 7$ |
| Reflections collected/unique | 895/250 [$R_{int.}= 0.0311$] |
| Completeness to 2theta | 95.8 % |
| Refinement method | Full-matrix least-squares on $F^2$ |
| Data/restraints/parameters | 250/0/10 |
| Goodness-of-fit on $F^2$ | 1.049 |
| Final R indices [I>2sigma(I)] | $R_1 = 0.0408$, w$R_2 = 0.0800$ |
| R indices (all data) | $R_1 = 0.0665$, w$R_2 = 0.0833$ |
| $\Delta\rho_{max}$ and $\Delta\rho_{min}$,(e/A$^3$) | 3.160 and -0.957 |

**Table 2.**

Atomic coordinates and equivalent isotropic and anisotropic displacement parameters [Å$^2 \times 10^3$] for the Li$_{0.95}$AsFe. $U_{iso}$ is defined as one third of the trace of the orthogonalized $U_{ij}$ tensor. The anisotropic displacement factor exponent takes the form: $-2\pi^2 [ (h^2a^2U_{11} + ... + 2hka^*b^*U_{12}]$. For symmetry reasons $U_{23}=U_{13}=U_{12}=0$.

| Atom | Site | x | y | z | $U_{iso}$ | $U_{11}=U_{22}$ | $U_{33}$ |
|---|---|---|---|---|---|---|---|
| As | 2c | 1/4 | 1/4 | 2625(1) | 19(1) | 16(1) | 24(1) |
| Fe | 2b | 3/4 | 1/4 | 1/2 | 18(1) | 16(1) | 22(1) |
| Li | 2a | 1/4 | 1/4 | 8480(30) | 27(6) | 17(6) | 45(12) |

The temperature dependence of the volume susceptibility of LiFeAs crystals, measured using both zero-field cooling (ZFC) and field cooling (FC) modes in a magnetic field of 1 mT applied along the *c*-axis, is shown in Figure 3. The transition temperature $T_c$ deduced from these curves is about 16 K, slightly lower than previously reported for stoichiometric single crystalline and polycrystalline samples [5,6]. One may note that the superconducting transition is rather broad, which could be due to disorder and/or to Li deficiency. The diamagnetic shielding value (ZFC), after a sample demagnetization correction, approaches 100% of -1, whereas the Meissner volume fraction is less that 10% of the ZFC value. This means that even in a field as low as 1 mT the magnetic flux expulsion is incomplete, with most of it being pinned and remaining trapped during the cooling process.

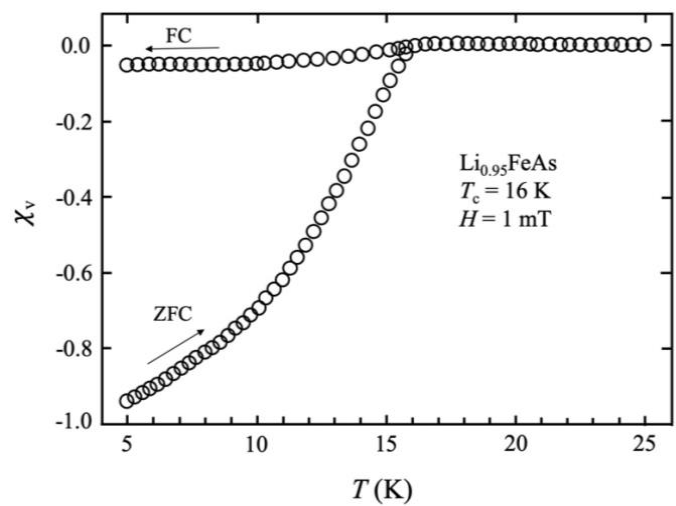

**Fig. 3.** Temperature-dependent volume magnetic susceptibility for a $Li_{0.95}FeAs$ single crystal measured in a field of 1 mT applied along the *c*-axis, after cooling in a zero field. ZFC and FC indicate zero-field cooling and field-cooling curves, respectively. The superconducting transition temperature $T_c$ was determined as the intersection point between the steepest tangent to a sigma (*T*) curve and the linear extension of the high-temperature of the curve.

The reduction in $T_c$ of our $Li_{0.95}FeAs$ crystals is relatively small, suggesting that the underlying mechanism of superconductivity in this material may be relatively insensitive to changes in stoichiometry within certain (~5%) limits. Obtaining higher levels of Li deficiency in LiFeAs is somewhat difficult and can depend on various factors, including synthesis methods, doping strategies, starting materials, and the tolerance of the material to deviations from stoichiometry. We attempted to use $Li_{0.6}FeAs$ as a starting composition and apply a pressure of 3 GPa to obtain crystals with a larger Li deficiency, but the outcome was unsuccessful. Li-deficient LiFeAs polycrystalline samples, with a relatively high level of deficiency, have occasionally been reported in the literature [9]. However, the exact level of Li deficiency in these samples is unknown and the superconducting properties may be significantly affected by other factors, such as the formation of secondary phases or other undesirable effects.

Polycrystalline Li$_{1-x}$FeAs ($x$ = 0, 0.2, 0.4) samples synthesized under high pressure conditions showed $T_c$ ranging from 16 K to 18 K [9]. The almost constant superconducting transition temperature implies the point superconducting feature of Li$_{1\pm x}$FeAs [5], in contrast with the high-$T_c$ cuprates and 1111 and 122 iron-based superconducting families, where the $T_c$ usually evolves systematically with the carrier density. The magnetic susceptibility of samples with a nominal composition Li$_{0.6}$FeAs in [9] shows a rather high $T_c$ of about 16 K, but with a substantially reduced superconducting volume and a positive background in the normal state. This may indicate that the actual Li concentration is higher than 0.6 and that FeAs or other related compounds precipitate as impurities. Nevertheless, even considering that Li$_{0.6}$FeAs may be a multiphase material, the high $T_c$ appears surprising because in most superconductors, foreign phases tend to lower the $T_c$ value. Overall, based on our observations, as well as on published results [5,50-53], we conclude that Li-deficient samples are stable up to about 5% of total Li deficiency and higher Li-deficient samples are unstable both under normal- and high-pressure conditions. This explains why, when starting from supposedly high-deficient compositions, like Li$_{0.6}$FeAs, for example, significant amounts of impurities arise. However, the synthesis at higher pressures (> 6 GPa) could be intriguing to further understand this matter.

There is a series of reasons why $T_c$ does not change drastically for Li deficient LiFeAs. While the precise mechanism of superconductivity in Fe-based superconductors is still under investigation, it is generally understood that the electron pairing is mediated by spin fluctuations or other electronic interactions [55,56]. It seems that, in the LiFeAs case, the electron interaction is robust against deviations from stoichiometry, i.e., as long as the electronic structure and the key interactions mediating superconductivity remain relatively unchanged, the effect of Li deficiency on $T_c$ remains relatively small, at least, within a certain range of deviation (around ~ 5%) from ideal stoichiometry. Beyond this range, other effects, such as phase separation or formation of secondary phases, may become dominant, leading to more significant changes in $T_c$. This observation is supported by first-principles calculations showing only small changes in the density of states (DOS) at Fermi level with the Li content [57]. Angle-resolved photoemission spectroscopy (ARPES) measurements also show that a few percent of Li-deficiencies in Li$_{1-x}$FeAs does not dramatically change the hole- and electron Fermi pocket sizes and alter the Fermi surface nesting conditions [51].

What accounts for the significant differences in superconducting properties observed in LiFeAs samples that are structurally identical and nearly stoichiometric based on diffraction measurements? To examine this issue in greater detail, we performed a comparative analysis of the LiFeAs [5,6,11,23], Li$_{1-x}$FeAs, and Li$_{1-y}$Fe$_{1+y}$As systems [52], which revealed that, in fact, superconductivity is very sensitive to derivations from stoichiometry.

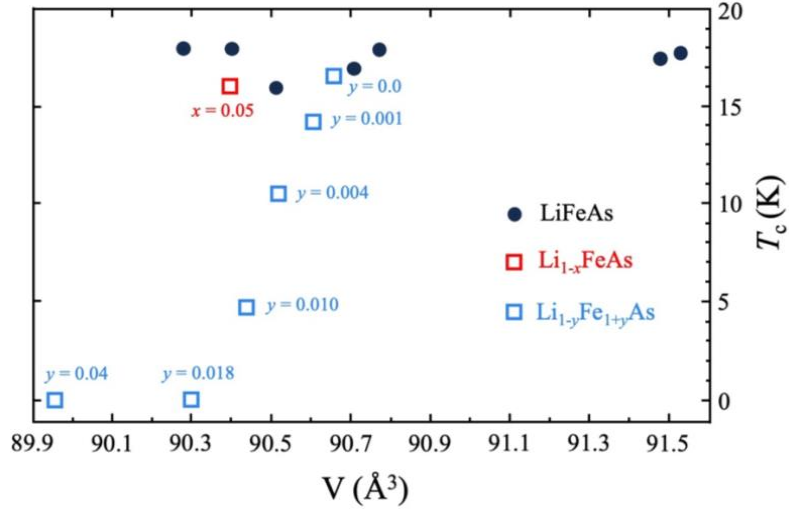

**Fig. 4.** The evolution of $T_c$ with unit cell volume for LiFeAs, $Li_{1-x}FeAs$, and $Li_{1-y}Fe_{1+y}As$ samples. The data for $Li_{1-x}FeAs$ system belong to present work. Data for LiFeAs and $Li_{1-y}Fe_{1+y}As$ were taken from Refs. [5,6,11,23] and [52], respectively.

In Figure 4, we show the evolution of $T_c$ vs the unit-cell volume for LiFeAs, $Li_{1-x}FeAs$, and $Li_{1-y}Fe_{1+y}As$ systems. First, we note that for stoichiometric and 5% deficient samples there is a significant variation in unit cell volumes, while $T_c$'s fall mostly within the 16-18 K range. Therefore, in LiFeAs and $Li_{1-x}FeAs$ systems, lattice compression has little effect on $T_c$ variation. The behavior of the $Li_{1-y}Fe_{1+y}As$ system is entirely different. Its superconducting transition temperature decreases linearly with the decreasing of the unit-cell volume down to ~90.42 Å, below which the system no longer supports superconductivity. Small over-stoichiometry in Fe leads to complete suppression in superconductivity. The extreme sensitivity of superconductivity in LiFeAs to Fe stoichiometry is quantitatively similar to that in the $Fe_{1+x}Se$ system. $Fe_{1.01}Se$ is a superconductor, while the incorporation of just 2% of additional Fe forms $Fe_{1.03}Se$ which breaks the superconducting state [58]. It is suggested that the magnetic pair-breaking by the Fe interstitial atoms is responsible for the suppression of the superconducting state in $Fe_{1+y}Se$. Analogously, we expect that a similar effect causes the same behavior in Fe-rich $Li_{1-y}Fe_{1+y}As$.

## 4. Summary and conclusions

By using a cubic-anvil high-pressure apparatus, single crystals of the LiFeAs superconductor were grown from a self-flux under a high pressure of 1 GPa at a temperature of 1050 °C. Single crystal X-ray diffraction confirmed the high quality of the grown crystals and showed a deficiency of 5% on the Li site. Temperature dependent magnetic susceptibility measurements revealed the occurrence of bulk superconductivity, with a superconducting transition at 16 K. Through comparison of LiFeAs with its nonstoichiometric derivatives, $Li_{1-x}FeAs$ and $Li_{1-y}Fe_{1+y}As$, it was demonstrated that superconductivity is less susceptible to Li deficiency, although it may disappear entirely in presence of a Fe excess. This observation reflects the behavior of the nonstoichiometric iron chalcogenide superconductor $Fe_{1+y}Se$ and may be due to the magnetic pair-breaking effect of the large local magnetic moments very close to the superconducting layers. When comparing the structural or physical properties of LiFeAs produced by different methods, one should be aware that non-stoichiometric derivatives of the $Li_{1-x}FeAs$ and $Li_{1-y}Fe_{1+y}As$ types may emerge during the synthesis, therefore, a careful control of composition is vital. Since the stoichiometry of LiFeAs is an important issue, single crystals with distinct Li and Fe stoichiometry are necessary for a comprehensive and accurate investigation of their basic characteristics. This will enhance our understanding of the mechanism of superconductivity in Fe-based superconductors.

**Acknowledgements**

The author acknowledges support from the Department of Earth and Planetary Sciences ETH Zurich and the Department of Chemistry, Biochemistry and Pharmaceutical Sciences of the University of Bern. The author would like to thank S. Katrych for help with the structural characterization and T. Shiroka for the critical reading of the manuscript and useful suggestions.

**References**

1. Y. Kamihara, T. Watanabe, M. Hirano, H. Hosono, *Iron-based layered superconductor $La(O_{1-x}F_x)FeAs$ (x = 0.05-0.12) with $T_c$ = 26 K*, J. Am. Chem. Soc. 130 (2008) 3296-3297. doi
2. Z.-A. Ren, W. Lu, J. Jang, W. Yi, X.-L. Shen, Z.-C. Li, G.-C. Che, X.-L. Dong, L.-L. Sun, F. Zhou, Z.-X. Zhao, *Superconductivity at 55 K in iron-based F-doped layered quaternary compound $Sm(O_{1-x}F_x)FeAs$*, Chin. Phys. Lett. 25 (2008) 2215-2216. doi
3. M. Rotter, M. Tegel, D. Johrendt, *Superconductivity at 38 K in the iron arsenide $(Ba_{1-x}K_x)Fe_2As_2$*, Phys. Rev. Lett. 101 (2008) 107006. doi
4. F.-C. Hsu, J.-Y. Luo, K.-W. Yeh, T.-K. Chen, T.-W. Huang, P.M. Wu, Y.-C. Lee, Y.-L. Huang, Y.-Y. Chu, D.-C. Yan, M.-K. Wu, *Superconductivity in the PbO-type structure α-FeSe*, Proc. Natl. Acd. Sci. USA 105 (2008) 14262-14264. doi
5. B. Lee, S. Khim, J.S. Kim, G.R. Stewart, K.H. Kim, *Single-crystal growth and superconducting properties of LiFeAs*, Eur. Phys. Lett. 91 (2010) 67002. doi
6. J.H. Tapp, Z. Tang, B. Lv, K. Sasmal, B. Lorenz, P.C.W. Chu, A.M. Guloy, *LiFeAs: An intrinsic FeAs-based superconductor with $T_c$ = 18 K*, Phys. Rev. B 78 (2008) 060505(R). doi
7. M.J. Pitcher, D.R. Parker, P. Adamson, S.J.C. Herkelrath, A.T. Boothroyd, R.M. Ibberson, M. Brunelli, S.J. Clarke, *Structure and superconductivity of LiFeAs*, Chem. Commun. 45 (2008) 5918-5920. doi
8. D.R. Parker, M.J. Pitcher, P.J. Baker, I. Franke, T. Lancaster, S.J. Blundell, S.J. Clarke, *Structure, antiferromagnetism and superconductivity of the layered iron arsenide NaFeAs*, Chem. Commun. 16 (2009) 2189-2191. doi
9. X.C. Wang, Q.Q. Liu, Y.X. Lv, W.B. Gao, L.X. Yang, R.C. Yu, F.Y. Li, C.Q. Jin, *The superconductivity at 18 K in LiFeAs system*, Solid State Commun. 148 (2008) 538-540. doi
10. Von R. Juza, K. Langer, *Ternäre phosphide und arsenide des lithium mit eisen, kobalt oder chrom im $Cu_2Sb$-typ*, Z. Anorg. Allg. Chem. 361 (1968) 58-73. doi
11. S. Chi, S. Grothe, R. Liang, P. Dosanjh, W.N. Hardy, S.A. Burke, D.A. Bonn, Y. Pennec, *Scanning tunneling spectroscopy of superconducting LiFeAs single crystals: evidence for two nodeless energy gaps and coupling to a bosonic mode*, Phys. Rev. Lett. 109 (2012) 087002. doi
12. L. Cao, W. Liu, G. Li, G. Dai, Qi Zheng, Y. Wang, K. Jiang, S. Zhu, Li Huang, L. Kong, et al., *Two distinct superconducting states controlled by orientations of local wrinkles in LiFeAs*, Nat. Commun. 12 (2021) 6312. doi

13. Y. Su, P. Link, A. Schneidewind, T. Wolf, P. Adelmann, Y. Xiao, M. Meven, R. Mittal, M. Rotter, D. Johrendt, et al. *Antiferromagnetic ordering and structural phase transition in BaFe$_2$As$_2$ with Sn incorporated from the growth flux*, Phys. Rev. B 79 (2009) 064504. doi
14. N.Ni, S.L. Budko, A. Kreyssig, S. Nandi, G.E. Rustan, A.I. Goldman, S. Gupta, J.D. Corbett, A. Kracher, P.C. Canfield, *Anisotropic thermodynamic and transport properties of single-crystalline Ba$_{1-x}$K$_x$Fe$_2$As$_2$ (x = 0 and 0.45)*, Phys. Rev. B 78 (2008) 014507. doi
15. J.S. Kim, S. Khim, L. Yan, N. Manivannan, Y. Liu, I. Kim, G.R. Stewart, K.H. Kim, *Evidence for coexistence of superconductivity and magnetism in single crystals of Co-doped SrFe$_2$As$_2$*, J. Phys.: Condens. Matter 21 (2009) 102203. doi
16. E. Dengler, J. Delsenhofer, H.-A. Krug von Nidda, S. Khim, J.S. Kim, K.H. Kim, F. Casper, C. Felser, A. Loidl, *Strong reduction of the Korringa relaxation in the spin-density wave regime of EuFe$_2$As$_2$ observed by electron spin resonance*, Phys. Rev. B 81 (2020) 024406. doi
17. N. Harrison, R.D. McDonald, C.H. Mielke, E.D. Bauer, F. Ronning, J.D. Thompson, *Quantum oscillations in antiferromagnetic CaFe$_2$As$_2$ on the brink of superconductivity*, J. Phys.: Condens. Matter 21 (2009) 322202. doi
18. Z. Bukowski, S. Weyeneth, R. Puzniak, P. Moll, S. Katrych, N.D. Zhigadlo, J. Karpinski, B. Batlogg, *Superconductivity at 23 K and low anisotropy in Rb-substituted BaFe$_2$As$_2$ single crystals*, Phys. Rev. B 79 (2009) 104521. doi
19. C.W. Chu, F. Chen, M. Gooch, A.M. Guloy, B. Lorenz, B. Lv, K. Sasmal, Z.J. Tang, J.H. Tapp, Y.Y. Xue, *The synthesis and characterization of LiFeAs and NaFeAs*, Physica C 469 (2009) 326-331. doi
20. Y.J. Song, J.S. Ghim, B.H. Min, Y.S. Kwon, M.H. Jung, J.S. Rhyee, *Synthesis, anisotropy, and superconducting properties of LiFeAs single crystal*, Appl. Phys. Lett. 96 (2010) 212508. doi
21. S. Kasahara, K. Hashimoto, H. Ikeda, T. Terashima, Y. Matsuda, T. Shibauchi, *Contrasts in electron correlations and inelastic scattering between LiFeP and LiFeAs revealed by charge transport*, Phys. Rev. B 85 (2012) 060503(R). doi
22. X.C. Wang, Q.Q. Liu, Y.X. Lv, Z. Deng, K. Zhao, R.C. Yu, J.L. Zhu, C.Q. Jin, Superconducting properties of "111" type LiFeAs iron arsenide single crystals, Sci. China-Phys. Mech. Ast. 53 (2010) 1199-1201. doi
23. I. Morozov, A. Boltalin, O. Volkova, A. Vasiliev, O. Kataeva, U. Stockert, M. Abdel-Hafiez, D. Bombor, A. Bachmann, L. Harnagea, et al., *Single crystal growth and characterization of superconducting LiFeAs*, Cryst. Growth Des. 10 (2010) 4428-4432. doi
24. N.D. Zhigadlo, S. Katrych, Z. Bukowski, S. Weyeneth, R. Puzniak, J. Karpisnki, *Single crystals of superconducting SmFeAsO$_{1-x}$F$_y$ grown at high pressure*, J. Phys.: Condens. Matter 20 (2008) 342202. doi
25. N.D. Zhigadlo, S. Katrych, S. Weyeneth, R. Puzniak, P.J.W. Moll, Z. Bukowski, J. Karpinski, H. Keller, B. Batlogg, *Th-substituted SmFeAsO: structural details and superconductivity with T$_c$ above 50 K*, Phys. Rev. B 82 (2010) 064517. doi
26. N.D. Zhigadlo, S. Katrych, M. Bendele, P.J.W. Moll, M. Tortello, S. Weyeneth, V.Yu. Pomjakushin, J. Kanter, R. Puzniak, Z. Bukowski, et al., *Interplay of composition, structure, magnetism, and superconductivity in SmFeAs$_{1-x}$P$_x$O$_{1-y}$*, Phys. Rev. B 84 (2011) 134526. doi
27. N.D. Zhigadlo, S. Weyeneth, S. Katrych, P.J.W. Moll, K. Rogacki, S. Bosma, R. Puzniak, J. Karpinski, B. Batlogg, *High-pressure flux growth, structural, and superconducting properties of LnFeAsO (Ln = Pr, Nd, Sm) single crystals*, Phys. Rev. B 86 (2012) 214509. doi
28. N.D. Zhigadlo, R. Puzniak, P.J.W. Moll, F. Bernardini, T. Shiroka, *Emergence of superconductivity in single-crystalline LaFeAsO under simultaneous Sm and P substitution*, J. Alloys Compd. 958 (2023) 170384. doi
29. N.D. Zhigadlo, *Growth of whisker-like and bulk single crystals of PrFeAs(O,F) under high pressure*, J. Cryst. Growth 382 (2013) 75-79. doi
30. N.D. Zhigadlo, *High pressure crystal growth of the antiperovskite centrosymmetric superconductor SrPt$_3$P*, J. Cryst. Growth 455 (2016) 94-98. doi
31. N.D. Zhigadlo, *Crystal growth and characterization of the antiperovskite superconductor MgC$_{1-x}$Ni$_{3-y}$*, J. Cryst. Growth 520 (2019) 56-61. doi
32. N.D. Zhigadlo, D. Logvinovich, V.A. Stepanov, R.S. Gonnelli, D. Daghero, *Crystal growth, characterization and advanced study of the noncentrosymmetric superconductor Mo$_3$Al$_2$C*, Phys. Rev. B 97 (2018) 214518. doi
33. R. Khasanov, Z. Guguchia, I. Eremin, H. Luetkens, A. Amato, P.K. Biswas, C. Rüegg, M.A. Susner, A.S. Sefat, N.D. Zhigadlo, et al., *Pressure-induced electronic phase separation of magnetism and superconductivity in CrAs*, Sci. Reports. 5 (2015) 13788. doi
34. N.D. Zhigadlo, N. Barbero, T. Shiroka, *Growth of bulk single-crystal MnP helimagnet and its structural and NMR characterization*, J. Alloys Compd. 725 (2017) 1027-1034. doi
35. N.D. Zhigadlo, *High-pressure growth and characterization of bulk MnAs single crystals*, J. Cryst. Growth 480 (2017) 148-153. doi

36. N.D. Zhigadlo, *Spontaneous growth of diamond from MnNi solvent-catalyst using opposed anvil-type high-pressure apparatus*, J. Cryst. Growth 395 (2014) 1-4. doi
37. N.D. Zhigadlo, J. Karpinski, *High-pressure synthesis and superconductivity of $Ca_{2-x}Na_xCuO_2Cl_2$*, Physica C 460-462 (2007) 372-373. doi
38. R. Khasanov, N.D. Zhigadlo, J. Karpinski, H. Keller, *In-plane magnetic penetration depth $\lambda_{ab}$ in $Ca_{2-x}Na_xCuO_2Cl_2$: Role of the apical sites*, Phys. Rev. B 76 (2007) 094505. doi
39. G. Schuck, S.M. Kazakov, K. Rogacki, N.D. Zhigadlo, J. Karpinski, *Crystal growth, structure, and superconducting properties of the b-pyrochlore $KOs_2O_6$*, Phys. Rev. B 73 (2006) 144506. doi
40. N.D. Zhigadlo, *High-pressure hydrothermal growth and characterization of $Sr_3Os_4O_{14}$ single crystals*, J. Cryst. Growth 623 (2023) 127407. doi
41. A. Ricci, N. Poccia, B. Joseph, L. Barba, G. Arrighetti, G. Ciasca, J.-Q. Yan, R.W. McCallum, T.A. Lograsso, N.D. Zhigadlo, et al., *Structural phase transition and superlattice misfit strain of RFeAsO (R = La, Pr, Nd, Sm)*, Phys. Rev. B 82 (2010) 144507. doi
42. D. Daghero, E. Piatti, N.D. Zhigadlo, G.A. Ummarino, N. Barbero, T. Shiroka, *Superconductivity of underdoped PrFeAs(O,F) investigated via point-contact spectroscopy and nuclear magnetic resonance*, Phys. Rev. B 102 (2020) 104513. doi
43. C. Müller, N.D. Zhigadlo, A. Kumar, M.A. Baklar, J. Karpinski, P. Smith, T. Kreouzis, N. Stingelin, *Enhanced charge-carrier mobility in high-pressure-crystallized poly(3-hexylthiophene)*, Macromolecules 44 (2011) 1221-1225. doi
44. N.D. Zhigadlo, *Crystal growth of hexagonal boron nitride (hBN) from Mg-B-N solvent system under high pressure*, J. Cryst. Growth 402 (2014) 308-311. doi
45. P.K. Sahoo, S. Memaran, F.A. Nugera, Y. Xin, T.D. Marquez, Z. Lu, W. Zheng, N.D. Zhigadlo, D. Smirnov, L. Balicas, et al., *Bilayer lateral heterostructures of transition metal dichalcogenides and their optoelectronic response*, ACS Nano 13 (2019) 12372-12384. doi
46. L. Khalil, C. Ernandes, J. Avila, A. Rousseau, P. Dudin, N.D. Zhigadlo, G. Cassabois, B. Gil, F. Oehler, J. Chaste, et al., *High p doped and robust band structure in Mg-doped hexagonal boron nitride*, Nanoscale Adv. 5 (2023) 3225-3232. doi
47. N.D. Zhigadlo, *Exploring 2D materials by high pressure synthesis: hBN, Mg-hBN, b-P, b-AsP, and GeAs*, J. Cryst. Growth 631 (2024) 127627. doi
48. G. Sheldrick, SHELXS-97, *Program for the Solution of Crystal Structures* (University of Göttingen, Germany, 1997); LXL-97, *Program for the Refinement of Crystal Structures* (University of Göttingen, Germany, 1997). doi
49. A. Seitkan, G.I. Lampronti, R.N. Widmer, N.P.M. Casati, S.A.T. Redfern, *Thermal behavior of iron arsenides under non-oxidizing conditions*, ACS Omega 5 (2020) 6423-6428. doi
50. S.J. Zhang, X.C. Wang, R. Sammynaiken, J.S. Tse, L.X. Yang, Z. Li, Q.Q. Liu, S. Desgreniers, Y. Yao, H.Z. Liu, et al., *Effect of pressure on iron arsenide superconductor $Li_xFeAs$ (x = 0.8, 1.0, 1.1)*, Phys. Rev. B 80 (2009) 014506. doi
51. M. Wang, M. Wang, H. Miao, S.V. Carr, D.L. Abernathy, M.B. Stone, X.C. Wang, L. Xing, C.Q. Jin, X. Zhang, et al, *Effect of Li-deficiency impurities on the electron-overdoped LiFeAs superconductor*, Phys. Rev. B 86 (2012) 144511. doi
52. M.J. Pitcher, T. Lancaster, J.D. Wright, I. Franke, A.J. Steele, P.J. Baker, F.L. Pratt, W.T. Thomas, D.R. Parker, S.J. Blundell, S.J. Clarke, *Compositional control of the superconducting properties of LiFeAs*, J. Am. Chem. Soc. 132 (2010) 10467-10476. doi
53. M. Wang, X.C. Wang, D.L. Abernathy, L.W. Harriger, H.Q. Luo, Y. Zhao, J.W. Lynn, Q.Q. Liu, C.Q. Jin, C. Fang, et al., *Antiferromagnetic spin excitations in single crystals of nonsuperconducting $Li_{1-x}FeAs$*, Phys. Rev. B 83 (2011) 220515(R). doi
54. W. Han, X.C. Wang, J.J. Gu, Q.Q. Liu, Z. Deng, C.Q. Jin, *Doping effect on the physical properties of $Li_xFe_{2-x}As$*, Int. J. Mod. Phys. B 29 (2015) 1550019. doi
55. S.V. Borisenko, D.V. Evtushinsky, Z.-H. Liu, I. Morozov, R. Kappenberger, S. Wurmehl, B. Büchner, A.N. Yaresko, T.K. Kim, M. Hoesch, et al., *Direct observation of spin-orbit coupling in iron-based superconductors*, Nat. Phys. 12 (2016) 311-317. doi
56. D.V. Evtushinsky, A.N. Yaresko, V.B. Zabolotnyy, J. Maletz, T.K. Kim, A.A. Kordyuk, M.S. Viazovska, M. Roslova, I. Morozov, R. Beck, et al., *High-energy electronic interaction in the 3d band of high-temperature iron-based superconductors*, Phys. Rev. B 96 (2017) 060501(R). doi
57. D.J. Singh, *Electronic structure and doping in $BaFe_2As_2$ and LiFeAs: density functional calculations*, Phys. Rev. B 78 (2008) 094511. doi
58. T.M. McQueen, Q. Huang, V. Ksenofontov, C. Felser, Q. Xu, H. Zandbergen, Y.S. Hor, J. Allred, A.J. Williams, D. Qu, et al., *Extreme sensitivity of superconductivity to stoichiometry in $Fe_{1+\delta}Se$*, Phys. Rev. B 79 (2009) 014522. doi